\documentclass[12pt,preprintnumbers,preprint,amsmath,amssymb,floatfix,prb]{revtex4}

\usepackage{graphicx}
\usepackage{dcolumn}
\usepackage{bm}
\usepackage{longtable}
\usepackage{graphics}
\usepackage{xspace}
\usepackage{epstopdf}
\usepackage{epsfig}
\newcommand{\um}{$U_m$ }
\newcommand{\umn}{$U_m$}
\newcommand{\umv}{$U_m^{(v)}$ }
\newcommand{\umvn}{$U_m^{(v)}$}
\newcommand{\umvc}{$U_{m c}^{(v)}$ }
\newcommand{\umvcn}{$U_{m c}^{(v)}$}

\newcommand {\ibid}{{\it ibid}. }
\newcommand {\etal}{{\it et al}. }

\renewcommand {\vec}{\mathbf}

\newcommand {\ET}{ET }

\newcommand {\bI}{$\beta$-ET$_2$I$_3$ }
\newcommand {\bIn}{$\beta$-ET$_2$I$_3$}

\newcommand {\bH}{$\beta_\textrm{H}$ }
\newcommand {\bL}{$\beta_\textrm{L}$ }

\newcommand {\kCl}{$\kappa$-ET$_2$Cu[N(CN)$_2$]Cl }

\begin{document}
\title{
Towards the parameterisation of the Hubbard model for salts of BEDT-TTF: A density functional study of isolated molecules.}
\author{Edan Scriven}
\email{edan@physics.uq.edu.au}
\author{B. J. Powell}
\affiliation{Centre for Organic and Photonic Electronics, School of Physical Sciences, University of Queensland, Qld 4072, Australia}

\begin{abstract}
We calculate the effective Coulomb repulsion between electrons/holes, \umvn, and site energy for an isolated BEDT-TTF [bis(ethylenedithio)tetrathiafulvalene] molecule {\it in vacuo}. $U_m^{(v)}=4.2 \pm0.1$ eV for 44 experimental geometries taken from a broad range of conformations, polymorphs, anions, temperatures, and pressures (the quoted `error' is one standard deviation). Hence we conclude that \umv is essentially the same for all of the compounds studied. This shows that the strong (hydrostatic and chemical) pressure dependence observed in the phase diagrams of the BEDT-TTF salts is not due to \umvn. Therefore, if the Hubbard model is  sufficient to describe the phase diagram of the BEDT-TTF salts, there must be significant pressure dependence on the intramolecular terms in the Hamiltonian and/or the reduction of the Hubbard $U$ due to the interaction of the molecule with the polarisable crystal environment. The renormalised value of \umv is significantly smaller than the bare value of the Coulomb integral: $F_0=5.2\pm0.1$ eV across the same set of geometries, emphasising the importance of using the renormalised value of \umvn. The site energy (for holes), $\xi_m=5.0\pm0.2$ eV, varies only a little more than \umv across the same set of geometries. However, we argue that this variation in the site energy plays a key role in understanding the role of disorder in ET salts. We explain the differences between the $\beta_L$ and $\beta_H$ phases of (BEDT-TTF)$_2$I$_3$ on the basis of calculations of the effects of disorder.
\end{abstract}

\maketitle

\section{Introduction}

\begin{figure}
	\centering 		\epsfig{figure=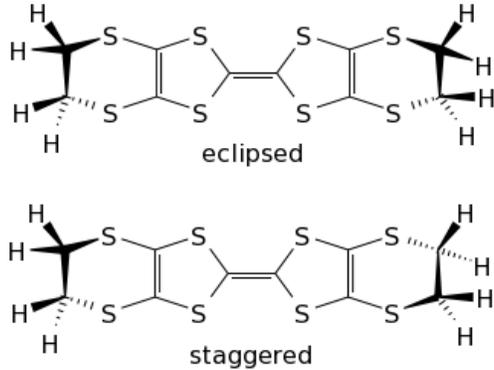, width=70mm, angle=0} 		\ 	\caption{The eclipsed and staggered conformations of the ET molecule.  } 	\label{fig:2D conformations}
\end{figure}

It is well known that density-functional theory (DFT), as implemented using the current generation of approximate exchange-correlation functionals, fails to capture the physics of strongly correlated electrons, for example, the Mott insulating state.\cite{YangScience} The electrons are strongly-correlated in the layered organic charge transfer salts of the form (ET)$_2X$, where ET is bis(ethylenedithio)tetrathiafulvalene (also known as BEDT-TTF and shown in Fig. \ref{fig:2D conformations}) and $X$ is a monovalent anion.\cite{JaimeRev,Powell:strongCorrelations} A great deal of interesting physics is driven by these strong electronic correlations including unconventional superconductor, Mott insulator, charge ordered insulator, bad metal, charge ordered metal, and pseudogap phases.\cite{JaimeRev,Powell:strongCorrelations} To date, first principles atomistic theory has not been able to describe the full range of physical phenomena observed in the ET salts. Therefore most theoretical approaches to describing organic charge transfer salts have been based on effective low-energy Hamiltonians, such as Hubbard models.\cite{JaimeRev,Powell:strongCorrelations} However, in order to make a detailed comparison between these calculations and experiment accurate parameterisations of the relevant effective low-energy Hamiltonians are required.

The strong electronic correlation effects evident in the macroscopic behaviour of the ET salts do not arise because the  intramolecular Coulomb interactions are unusually strong. Rather, the electrons move in narrow bands and thus the `kinetic' energy associated with electrons hopping between molecules is much smaller the equivalent energy scale in elemental metals; thus the narrow bands give rise to the strong correlations.\cite{JaimeRev,Powell:strongCorrelations} This means that density functional theory (DFT) is likely to give accurate results for single ET molecules, even if it cannot accurately calculate the properties of the extended system. Similar statements hold for the alkali doped fullerides,\cite{GunnarssonBook} oligo-acene and thiopenes,\cite{Brocks} and the organic conductor tetrathiafulvalene-tetracyanoquinodimethane (TTF-TCNQ)\cite{Laura} where significant insight has been gained from parameterising effective low-energy Hamiltonians from DFT calculations on a single (or a few) molecules. There have been some pioneering efforts\cite{other:Imamura,other:Fortunelli,other:Ducasse,other:Castet,vibronic} to apply this approach to ET salts, but no systematic study of the wide range of ET salts known to synthetic chemistry. Many other families of organic charge transfer salts based on other donor and, even, acceptor molecules are also known to show strong correlation effects.\cite{Ishiguro,chem-rev}

In this paper we present DFT calculations for a single ET molecule in vacuum. By studying various charge states we calculate parameters for the Hubbard model, specifically the effective Coulomb repulsion between two electrons/holes, `the Hubbard $U$', and the site energy. In section \ref{sect:model} we define the problem and stress the difference between the renormalised and bare values of the parameters we calculate below. In section \ref{sect:bench} we compare and contrast various methods of calculating the Hubbard parameters and provide benchmarks for our calculations against the previous literature. In section \ref{sect:main} we report our main results: the values of the parameters calculated at 44 distinct geometries observed by x-ray scattering in a broad range of conformations, polymorphs, anions, temperatures, and pressures. We find no significant variation in the Hubbard $U$, but a slightly larger variation in the site energy. We discuss the importance of this variation of the site energy for the effect of disorder on the superconducting state of the ET salts. In particular we are able to explain the origin of the difference between the superconducting critical temperatures of the $\beta_H$ and $\beta_L$ phase of (ET)$_2$I$_3$ on the basis of these results and the theory of non-magnetic impurity scattering in unconventional superconductors.
The technical details of our DFT calculations are reported in section \ref{sect:meth} and our conclusions are drawn in section \ref{sect:conc}.
We will not attempt to solve the Hubbard models relevant to the ET salts in this paper; this is an extremely demanding task which has been the subject of numerous previous studies.\cite{Powell:strongCorrelations,JaimeRev}

\section{The Hubbard model and density functional theory}\label{sect:model}

It has frequently been argued\cite{Powell:strongCorrelations,reviews,JaimeRev} that the physics of ET salts can be understood in terms of (various flavours of) Hubbard models. Here two different classes of ET salt need to be distinguished. In the first class, which includes the $\kappa$, $\beta$, and $\beta'$ phases, the ET molecules are strongly dimerised. The intra-dimer dynamics are typically integrated out of the effective low-energy Hamiltonian, to leave one with a half-filled Hubbard model, where each dimer is treated as a site.\cite{Powell:strongCorrelations} However, if one is to satisfactorily parameterise this effective Hamiltonian, one must first understand the intra-dimer dynamics. In the second class of ET salt, which includes the $\alpha$, $\beta''$, and $\theta$ phases, the dimerisation is weak or absent and one is forced to consider a quarter-filled Hubbard model in which each site corresponds to a single monomer.\cite{JaimeRev} 

It is convenient, for our purposes, to write the (extended) Hubbard model in the form
\begin{eqnarray}
\hat{\cal H}=\sum_i\hat{\cal H}_i+\sum_{ij}\hat{\cal H}_{ij},
\end{eqnarray}
where $\hat{\cal H}_i$ are the terms that depend on the physics of a single ET molecule and $\hat{\cal H}_{ij}$ are the terms coupling more than one ET molecule. In this paper we assume that
\begin{eqnarray}
\hat{\cal H}_i=E_0+\epsilon_m \hat c_{i\sigma}^\dagger \hat c_{i\sigma}+U_m \hat c_{i\uparrow} ^\dagger  \hat c_{i\uparrow} \hat c_{i\downarrow} ^\dagger \hat c_{i\downarrow}, \label{eqn:sshh}
\end{eqnarray}
where $\hat c_{i\sigma}^{(\dagger)}$ annihilates (creates) a particle with spin $\sigma$ on site $i$, $E_0$ may be thought of as the energy due to the `core electrons' and the nuclei [a more rigourous definition is given below Eq. (\ref{eqn:en})], $\epsilon_m$ is the site energy for site $i$, and $U_m$ is the effective Coulomb repulsion between electrons on site $i$, whose interpretation we will discuss presently. (In general $E_0$, $\epsilon_m$, and $U_m$ may be different of different sites, but we suppress site labels in order to simplify notation. The subscript $m$ serves to remind us that we are dealing with the Hubbard parameters for a monomer rather than dimeric parameters.)  Note that we do not consider vibronic interactions; however these can straightforwardly be included via the Hubbard-Holstein model. Calculations of the vibronic interactions have been reported by several groups.\cite{order1,vibronic} Also note that, we will not discuss the form or the parameterisation of $\hat{\cal H}_{ij}$ below. Most previous studies of $\hat{\cal H}_{ij}$ have been based on the extended H\"uckel
or tight binding approximation, which neglect all two-body terms in $\hat{\cal H}_{ij}$.\cite{Ishiguro,Canadell}

It is convenient to separate \um into two terms
\begin{eqnarray}
U_m=U_m^{(v)}-\delta U_m^{(p)},
\end{eqnarray}
where \umv is the value of \um for an isolated ET molecule in vacuum and $\delta U_m^{(p)}$ is the reduction of \um when the molecule is placed in  the polarisable crystalline environment. $\delta U_m^{(p)}$ has been successfully calculated for alkali doped fullerenes,\cite{Martin,Antropov,Quong}  oligo-acene and thiopenes,\cite{Brocks} and TTF-TCNQ.\cite{Laura} 
Perhaps the simplest system for which to calculate $\delta U_m^{(p)}$ is the alkali doped fullerides. This system has (approximate) spherical symmetry and a large intermolecular spacing relative to the size of the molecule. $\delta U_m^{(p)}$ has been calculated for the alkali doped fullerides both by assuming that the crystal is a dielectric medium and by assuming that each molecule in the crystal contributes to $\delta U_m^{(p)}$ only via the classical polarisability tensor appropriate to that molecule.\cite{Gunnarsson,Martin,Antropov,Quong} These calculations show that $\delta U_m^{(p)}$ can be a significant fraction of \umvn,  perhaps as much as a half.\cite{Gunnarsson,Martin,Antropov,Quong}
However, the calculation of the polarisation correction to the U for the ET salts is a much more difficult problem. First, the systems have very low symmetry. Second, many ET salts have polymeric counterions, therefore it is not possible to approximate the anion layers by a lattice of dipoles. Third, ET molecules are separated by distances typical of $\pi$-stacking in aromatic compounds ($\sim$3.5-4 \AA), i.e., distances that are quite small compared to the size of the molecule. This invalidates the approximations based on well-separated dipoles or an effective mean-field dielectric medium that have been successfully applied to the fullerenes. Therefore, we limit our attention on the calculation of \umv in this paper. Clearly the calculation of $\delta U_m^{(p)}$ is an important problem that must be solved before a full parameterisation of the Hubbard model for ET salts can be completed. Interestingly, Cano-Cort\'es \etal\cite{Laura} have recently reported a calculation of $\delta U_m^{(p)}$ for TTF-TCNQ, which manifests some of these problems, where the dipole moment is associated with individual atoms, rather than entire molecules.

It is often stated that the Hubbard $U$ is the Coulomb repulsion between two electrons in the highest occupied molecular orbital (HOMO) of an ET molecule.\cite{wrongU} This is, indeed, often a conceptually helpful way to think about the Hubbard $U$. However, if one were to take this literally it would be natural to equate \umv with the zeroth Slater-Condon parameter, i.e., the Coulomb integral,
\begin{equation}
F_0 = \int d^3 {\bm r}_1 \int d^3 {\bm r}_2  \frac{\rho_\uparrow({\bm r}_1) \rho_\downarrow({\bm r}_2)}{\left|{\bm r}_1-{\bm r}_2\right|}, \label{eqn:F0}
\end{equation}
where $\rho_\sigma({\bm r})$ the density of spin $\sigma$ electrons at the position $\bm r$ in the HOMO of the ET molecule. $F_0$ is rather different from the Hubbard $U$. To see this 
recall that the Hubbard model is an effective low-energy Hamiltonian and therefore the Hubbard parameters are renormalised. Assuming that $U_m^{(v)}=F_0$ is equivalent to simply ignoring the high energy degrees of freedom not contained in the Hubbard model. This is clearly inadequate: $F_0$ is the unrenormalised value of \umvn. In principle one should derive the effective low-energy Hamiltonian by explicitly integrating out the high energy degrees of freedom. However, this is not practicable  in this context for all but the simplest systems.\cite{Gunnarsson,FreedSimple} Therefore, one should construct the effective low-energy Hamiltonian so as to capture the relevant degrees of freedom and parameterise this effective Hamiltonian appropriately. We now review how to do this for the Hubbard model.

It is trivial to solve the single site Hubbard model (\ref{eqn:sshh}); one finds that
\begin{subequations}\label{eqn:en}
\begin{eqnarray}
E_1 &=& E_0 + \epsilon_m \\ 
E_2 &=& E_0 + 2 \epsilon_m + U_m
\end{eqnarray}
\end{subequations}
where $E_n$ is the energy of the model with $n$ electrons. Note that $E_n$ may be calculated in DFT as it is simply the ground state total energy of ET$^{(2-n)+}$ and DFT does well for total energies. It follow immediately from (\ref{eqn:en}) that
\begin{equation}
U_m = E_0 + E_2 - 2 E_1. \label{eqn:defnU}
\end{equation}
This can be visualised as the energy cost of the charge disproportionation reaction $2(\textrm{ET}^+)\rightarrow \textrm{ET}+\textrm{ET}^{2+}$ for infinitely separated ET molecules. Another interpretation is that \um is the charge gap, i.e., the difference between the chemical potentials for particles and holes, for ET$^+$. It is also clear from Eq. (\ref{eqn:en}) that the site energy is given by
\begin{equation}
\epsilon_m = E_1 -  E_0. \label{eqn:defnEpsilon}
\end{equation}
This is simply the second ionisation energy of the ET molecule, $\textrm{ET}^+\rightarrow \textrm{ET}^{2+}$.

 One can also write Eq.  (\ref{eqn:sshh}) in terms of the hole operators defined by $\hat h_{i\sigma}^\dagger\equiv\hat c_{i\sigma}$ which gives
\begin{eqnarray}
\hat{\cal H}_i=E_2+\xi_m \hat h_{i\sigma}^\dagger \hat h_{i\sigma}+U_m \hat h_{i\uparrow} ^\dagger  \hat h_{i\uparrow} \hat h_{i\downarrow} ^\dagger \hat h_{i\downarrow}, \label{eqn:sshhh}
\end{eqnarray}
where,
\begin{equation}
\xi_m = -(\epsilon_m+U_m)= E_1 -  E_2 \label{eqn:defnXi}
\end{equation}
is the site energy for holes. Physically $\xi_m$ corresponds to the ionisation energy of ET. Thus one sees that while the Hubbard $U$ is the same for electrons and holes, the site energy is not. Note that we limit our study to salts of the form ET$_2X$ where $X$ is a monovalent anion. In these materials the HOMO of the ET molecule is, on average, three quarters filled with electrons or, equivalently, one quarter filled with holes. Therefore, the hole parameters are of more physical relevance than the electron parameters, as we will discuss further below.

One might also consider the Taylor series for the dependence of the energy on the `classical' (i.e., not quantised) charge in terms of electrons,
\begin{subequations}
\begin{equation}
E(q) \simeq E(0) + \left.\frac{\partial E}{\partial q}\right|_{q=2^-} (2-q) + \frac{1}{2} \left.\frac{\partial^2E}{\partial q^2}\right|_{q=2^-} (2-q)^2,
\end{equation}
or holes,
\begin{equation}
E(q) \simeq E(2) + \left.\frac{\partial E}{\partial q}\right|_{q=0^+} q + \frac{1}{2} \left.\frac{\partial^2E}{\partial q^2}\right|_{q=0^+} q^2,
\end{equation}
\end{subequations}
where $q=2^-$ indicates the limit as $q\rightarrow2$ from below and $q=0^+$ indicates the limit as $q\rightarrow0$ from above. Again, $E(q)$ can be calculated from DFT by calculating the total energy of the relevant charge states. It then follows from (\ref{eqn:defnU}) that the classical $U$ is given by
\begin{eqnarray}
U_{mc}  =  \frac{\partial^2E}{\partial q^2}.
\end{eqnarray}
Similarly, the classical site energies are
\begin{eqnarray}
\epsilon_{mc} = \left.\frac{\partial E}{\partial q}\right|_{q=2^-}\textrm{ and}\hspace{10pt} \xi_{mc} = \left.\frac{\partial E}{\partial q}\right|_{q=0^+},
\end{eqnarray}
as one expects from Janak's theorem.\cite{Janak} We, therefore, are left with the classical results in terms of electrons,
\begin{subequations}\label{eqn:defnUf}
\begin{equation}
E(q) \simeq E(0) + \epsilon_{mc} (2-q) + \frac{1}{2} U_{mc} (2-q)^2. \label{eqn:defnUfe}
\end{equation}
and holes,
\begin{equation}
E(q) \simeq E(2) + \xi_{mc} q + \frac{1}{2} U_{mc} q^2. \label{eqn:defnUfh}
\end{equation}
\end{subequations}
Note, however, that $\xi_{mc} = -(\epsilon_{mc}+2U_{mc})$ in contrast to Eq. (\ref{eqn:defnXi}). This is a manifestation of the failure of classical approximation for the site energy.

The density functional formalism can be applied to fractional numbers of electrons.\cite{Janak,YangScience} Therefore one can calculate the energy for a large number of charge states and find the best fit values of $E(0)$, $\epsilon_{mc}$, $\xi_{mc}$, and $U_{mc}$ in Eqs. (\ref{eqn:defnUf}), which is operationally more satisfactory than determining the two free parameters ($\xi_m$ and $U_m$ determine $\epsilon_m $) from two data points [$E(1)-E(0)$ and $E(2)-E(0)$]. Of course this method suffers from the fact that, unlike Eqs. (\ref{eqn:defnU}), (\ref{eqn:defnEpsilon}), and (\ref{eqn:defnXi}),  Eqs. (\ref{eqn:defnUf}) are not exact.

\begin{figure}
	\centering 		\epsfig{figure=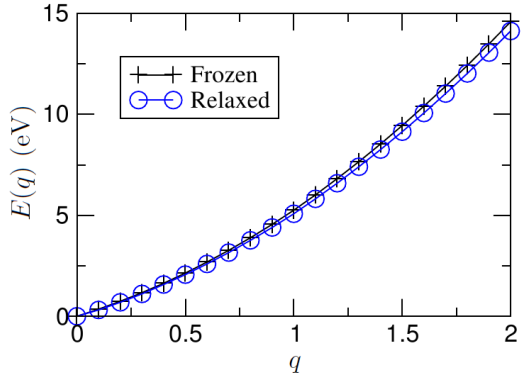, width=70mm, angle=0} 		\ 	\caption{(Color online) Dependence of the energy of an ET molecule on the charge of the molecule. Two different geometries are studied. The `frozen' geometry is relaxed from that found experimentally\cite{bI3:Leung} in $\beta$-(ET)$_2$I$_3$ in the neutral charge state and then held fixed during the self-consistent field (SCF) calculations at different charge states. For the `relaxed' data the nuclear geometry is relaxed separately for each charge state. The energies are exactly equal by definition in the charge neutral state; for other charge states the nuclear relaxation lowers energies. This reduces the curvature of $E(q)$, i.e., lowers \umvc and hence \umvn. It can be seen that for both the `frozen' and the `relaxed' geometries the energy of the fractional charge states is extremely well described by the classical quadratic functions, Eqs. (\ref{eqn:defnUf}). In contrast it is known that for the exact functional the energy of molecules with fractional charges is a linear interpolation between the integer charge states.\cite{YangScience,Janak,Perdew} We find that to an excellent approximation \umvcn=\umv whereas $\epsilon_{mc}$ and $\xi_{mc}$ are not good approximations to $\epsilon_m$ and $\xi_m$. In contrast for the exact functional $\epsilon_m=\epsilon_{mc}$, $\xi_m=\xi_{mc}$, and $U_{mc}^{(v)}=0$. This incorrect result is a manifestation of the delocalisation error of DFT and closely related to the band gap problem.\cite{YangScience,Yangsotherpapers,Perdew}  These calculations use  TZP basis sets and TM2 pseudopotentials.} 	\label{fig:frozen vs relaxed}
\end{figure}

 Eqs. (\ref{eqn:defnUf}) give remarkably good fits to the fractional charge data (c.f. Fig. \ref{fig:frozen vs relaxed}). Further we find that $U_{mc}^{(v)}$ is an extremely accurate approximation to \umv (typically the two values differ by $\lesssim1\%$). In contrast  $\epsilon_{mc}$ and $\xi_{mc}$ are not good approximations to $\epsilon_m$ and $\xi_m$. This is a consequence of our use of an approximate generalised gradient approximation (GGA)\cite{GGA} exchange-correlation functional. It is known that for the exact functional the energies of states with fractional charges are linear interpolations between integer charge states.\cite{Perdew,YangScience,Janak} In contrast for local functionals, such as the local density approximation (LDA) or that of Perdew, Burke and Erzenhof (PBE),\cite{PBE} $E(q)$ is a convex function.\cite{YangScience,Yangsotherpapers} This is closely related to the self-interaction error and leads to the delocalisation error in local functionals and the band gap problem.\cite{YangScience,Yangsotherpapers,Perdew} Indeed, for the exact functional $\epsilon_m=\epsilon_{mc}$, $\xi_m=\xi_{mc}$, and $U_{mc}^{(v)}=0$.\cite{Perdew,YangScience,Janak,Yangsotherpapers} 

\section{Computational methods}\label{sect:meth}

We used DFT to calculate the total energies and HOMO charge densities of ET molecules in various conformations and charge states. In order to compare the effects of different basis sets and pseudopotentials (including no pseudopotential) we made use of two different implementations of DFT, SIESTA\cite{SIESTA} and NRLMOL.\cite{NRLMOL,PP} In each case a separate self-consistent field (SCF) calculation was performed at every charge state studied.

SIESTA calculations were performed  using the PBE exchange-correlation functional.\cite{PBE} A triple-$\zeta$ plus single polarisation (TZP) basis set of Sankey numerical atomic orbitals\cite{Sankey} was used for all atoms. These orbitals are confined to some radius $r_c$ from their centres, which introduces a small increase to the energy of each orbital. The value of $r_c$ was determined by specifying the maximum allowed increase in energy due to this cutoff, which we limited to 2 mRy. The convergence of the integration mesh is determined by specifying an effective plane-wave energy cutoff for the mesh, which we set to 250 Ry. The initial spin moments on each atom were arranged antiferromagnetically, i.e., with opposite signs  on neighbouring atoms, wherever possible. All SIESTA calculations reported below used pseudopotentials  constructed according to the improved Troullier-Martins (TM2) method.\cite{TM2,this:sup}

NRLMOL performs massively parallel electronic structure calculation using Gaussian orbital methods. We again used the PBE exchange correlation functional. In the calculations presented in this paper we use Porezag-Pederson (PP) basis sets \cite{PP} which have been carefully optimized for the PBE-GGA energy functional using variational energy criteria. As discussed in Ref. \onlinecite{PP}, for each atom, the basis sets are optimized with respect to the total number of Gaussian decay parameters, and with respect to variation of these parameters and the contraction coefficients. These are roughly of triple to quadruple zeta quality.\cite{PP} As compared to other Gaussian basis sets, a key improvement in the PP optimization scheme is that the resulting basis sets satisfy what is now referred to as the $Z^{10/3}$ theorem. This theorem \cite{PP} discusses proper scaling of the Gaussian exponents near the nuclei as a function of atomic charge. It has been shown that the resulting PP basis sets exhibit no superposition error and alleviate the need for counterpoise corrections in weakly bound systems. We report both all electron calculations and calculations using pseudopotentials for C and S atoms performed in NRLMOL. These pseudopotentials were constructed using the method of Bachelet, Hamann and Schl\"{u}ter (BHS).\cite{BHS}

Nuclear positions for C and S atoms were obtained from x-ray crystallography.\cite{bI3:Leung,aI3:Bender,aI3:Emge,kI3:Kobayashi,tI3:Salameh,bAuI2:Wang,bIBr2:Williams,bI3:Madsen,bIAuBr:Ugawa,bAuBr2:Mori,kNCN2:Geiser,kCl3kbar:Schultz,kNCS2:Rahal,kNCS2:Schultz,kCN3:Geiser,kHg:Li,bSF5CH2CF5CO3:Geiser} In section \ref{sect:bench} we will discuss various schemes for relaxing these atomic positions and for calculating \umvn, therefore we delay defining these methods to that section.\cite{footgeo}  The total energy was then computed for charge states ranging from charge neutral (doubly-occupied HOMO), to the 2+ charge state (the same orbital unoccupied) in increments of 0.1 electrons. The resulting 21 data points per molecule were then fitted to a quadratic function by the method of least squares, from which values of \umvcn, $\epsilon_{m c}$ and $\xi_{m c}$ were extracted according to Eqs. (\ref{eqn:defnUf}). \umn, $\epsilon_m$ and $\xi_m$ were calculated from integer charge states according to Eqs. (\ref{eqn:defnU}), (\ref{eqn:defnEpsilon}), and  (\ref{eqn:defnXi}).

Calculations of the Coulomb integral, $F_0$, make use of the orbitals from the corresponding calculations. The integrals in Eq. (\ref{eqn:F0}) were calculated in the charge neutral state on a $55\times30\times20$ mesh using the trapezoidal rule. We have also investigated finer meshes and the calculations were found to be well converged with respect to the number of integration points. Equation (\ref{eqn:F0}) has a pole at $\vec{r}_1 = \vec{r}_2$. Corresponding terms in the numerical integral were approximated by the analytical solution of equation (\ref{eqn:F0}) for a sphere of uniform charge density and the same volume as the mesh volume elements. These points contribute $\sim2\%$ to the value of $F_0$, and the error associated with the uniform sphere approximation is estimated to be $\lesssim1\%$ of the contribution from individual mesh points, i.e. $\lesssim0.02\%$ of the total value of $F_0$.

\section{Geometry optimisation, pseudopotentials, basis sets, and calculation methods}\label{sect:bench}

We report results of calculations of \umv and $F_0$ performed using TZP basis sets and TM2 pseudopotentials in Table \ref{tab:bI3:differentMethods}. For these calculations we take the nuclear positions of the C and S atoms from x-ray scattering experiments\cite{bI3:Leung} on \bI and the nuclear positions of the H atoms were relaxed using the conjugate-gradient method.\cite{footgeo} Henceforth, we will refer to this method of determining the nuclear geometry as `experimental'. Both eclipsed and staggered conformations (c.f. Fig. \ref{fig:2D conformations}) are observed experimentally\cite{bI3:Leung} and we present results for both conformations. However, we find no significant differences between the eclipsed and staggered conformations in either \umv or $F_0$. We also find no significant difference in the calculated values of \umv and \umvcn. Similar results are found at the other geometries studied below.

\begin{table}
\begin{tabular}{|l|l|l|l|}
\hline
Quantity & Eclipsed & Staggered \\
\hline
\umvc & 4.12 & 4.07 \\
\umv & 4.17 & 4.08\\
$F_0$ & 5.40 & 5.35 \\
\hline
\end{tabular}
\caption{Comparison of different methods of calculating the effective monomer on-site Coulomb repulsion for ET. No significant differences are found between \umvc and \umv from the DFT calculations. This result has been confirmed for all of the geometries studied below. Further, no significant differences are found between the two conformations. $F_0$ is significantly larger than \umvn, in qualitative agreement with previous results from wavefunction based methods (c.f. Table \ref{tab:lit:otherUs}).  The nuclear geometry is that seen `experimentally' in $\beta$-(ET)$_2$I$_3$ for both the eclipsed and staggered conformations. The calculations use  TZP basis sets and TM2 pseudopotentials. All values are in eV.}\label{tab:bI3:differentMethods}
\end{table}

Several groups\cite{other:Imamura,other:Fortunelli,other:Ducasse,other:Castet} have reported calculated values of $F_0$ or \umv previously. We summarise their results in Table \ref{tab:lit:otherUs}. It can be seen from Tables \ref{tab:bI3:differentMethods} and \ref{tab:lit:otherUs} that there is broad agreement between the values calculated by a range of different levels of theory, basis sets, and starting from different experimental geometries. Across all of the calculations $F_0$ is $\sim$40\% larger than \umvn.  This over-estimation has been attributed to the fact that $F_0$ neglects orbital relaxation,\cite{other:Fortunelli,other:Ducasse,other:Castet} i.e., the tendency of the orbital to change as its population changes. Conceptually, this is very important. It stresses that the sites in the Hubbard model do not correspond to any particular `orbital' of the ET molecule, as these orbitals change in response to charge fluctuations, etc. in the crystal. $F_0$ is, in fact, simply the first order approximation to \umvn.\cite{FreedSecond,FreedBridge} Therefore the true value of \umv contains many non-trivial quantum many-body effects which are absent in $F_0$ beyond just orbital relaxation.\cite{footSIC,FreedSecond}

\begin{table}
\begin{tabular}{|l|l|l|l|}
\hline Method & Basis set & Atomic geometry from & $F_0$ \\ \hline HF\cite{other:Imamura} & SBK-31G*   & $\kappa$-(ET)$_2$Cu[N(CN)$_2$]Cl\cite{other:Imamura:source} & 5.90 \\
HF\cite{other:Imamura} & SBK-31G*   & $\kappa$-(ET)$_2$Cu[N(CN)$_2$]Br\cite{other:Imamura:source} & 5.90 \\
HF\cite{other:Imamura} & SBK-31G*   & $\kappa$-(ET)$_2$Cu[N(CN)$_2$]I\cite{other:Imamura:source} & 5.83 \\ 
RHF ($q=4$)\cite{other:Fortunelli} & 6-31G** &  $\kappa$-(ET)$_2$Cu[N(CN)$_2$]Br\cite{other:Fortunelli:source} & 5.44 \\
RHF ($q=2$)\cite{other:Fortunelli} & 6-31G** & $\kappa$-(ET)$_2$Cu[N(CN)$_2$]Br\cite{other:Fortunelli:source} & 6.40 \\
RHF ($q=0$)\cite{other:Fortunelli} & 6-31G**  & $\kappa$-(ET)$_2$Cu[N(CN)$_2$]Br\cite{other:Fortunelli:source} & 5.58 \\
\hline Mean $F_0$ & & & 5.84  \\ Std. dev. & & & 0.33 \\
\hline 
\hline Method & Basis set & Atomic geometry from & \umv \\
\hline RHF\cite{other:Fortunelli} & 6-31G**  & $\kappa$-(ET)$_2$Cu[N(CN)$_2$]Br\cite{other:Fortunelli:source} & 4.48 \\
VB-HF\cite{other:Ducasse} & AM1 & $\kappa$-(ET)$_2$Cu(NCS)$_2$\cite{other:Ducasse:source} & 3.92 \\
VB-HF\cite{other:Ducasse} & AM1  & $\kappa$-(ET)$_2$Cu(NCS)$_2$\cite{other:Ducasse:source} & 3.83 \\
VB-RHF\cite{other:Castet} & AM1  & $\beta$-(ET)$_2$I$_3$\cite{other:Ducasse:source} & 3.90 \\
VB-RHF\cite{other:Castet} & DZ 4-31G  & $\beta$-(ET)$_2$I$_3$\cite{other:Ducasse:source} & 4.61 \\
\hline Mean \umv & &  & 4.15  \\ Std. dev. & &  & 0.37 \\
\hline
\end{tabular}
\caption{Previously reported calculations of the Coulomb energy ($F_0$ and \umvn) of ET. These calculations  were performed at various levels of theory, using various basis sets and with geometries taken from x-ray crystallography experiments on various different materials. $F_0$ is significantly larger than \umvn, consistent with our results. We make use of the following abbreviations in the above table: HF (Hartree-Fock); RHF (restricted Hartree-Fock); and VB (valence bond). Fortunelli and Painelli\cite{other:Fortunelli} calculated $F_0$ for \ET dimers in different charge states, $q$. All energies are in eV.}\label{tab:lit:otherUs}
\end{table}

Now we turn to investigate the effect of different calculation schemes on the Hubbard parameters. To that end, results are presented for values of \umv for ET molecules in geometries taken from $\beta$-(ET)$_2$I$_3$, calculated with two major modifications to the calculation. The first modification is to relax the nuclear coordinates. We have investigated three methods of determining the nuclear coordinates.\cite{footgeo} The first nuclear co-ordinate set is the `experimental' geometry defined above, i.e., the experimental positions are used for the C and S atoms, but the H atoms, which are not seen in x-ray scattering experiments, are relaxed. The second set was obtained by relaxing all the nuclear positions in the charge neutral state and holding that geometry fixed for all later SCF calculations at different charge states. This set is labelled `frozen'. The third set of nuclear co-ordinates was obtained by relaxing all nuclear positions in each charge state. This set is labelled `relaxed'. The second modification to the computational method is to perform the calculations with different basis sets and pseudopotentials.  

\begin{table*}
\begin{tabular}{|l|l|l|l|l|l|l|l|l|l|l|}
\hline
Coordinates & Conformation & Basis set & Pseudo. & \umv & $\epsilon_m$ & $\xi_m$ & \umvc & $\epsilon_{m c}$ & $\xi_{m c}$ & RMSE \\
\hline
experimental & staggered & PP & All & 3.90 & -9.15 & 5.25 & 3.87 & -11.07 & 3.33 & 0.010 \\
experimental & staggered & PP & BHS & 3.93 & -9.19 & 5.26 & 3.90 & -11.13 & 3.34 & 0.012 \\
experimental & staggered & TZP & TM2 & 4.08 & -9.28 & 5.21 & 4.07 & -11.31 & 3.17 & 0.012 \\
frozen & staggered & PP & All & 3.78 & -9.63 & 5.85 & 3.76 & -11.50 & 3.99 & 0.009 \\
frozen & staggered & PP & BHS & 3.81 & -9.68 & 5.87 & 3.78 & -11.57 & 4.00 & 0.010 \\
frozen & staggered & TZP & TM2 & 4.11 & -9.28 & 5.17 & 4.09 & -11.33 & 3.14 & 0.010 \\
relaxed & staggered & TZP & TM2 & 4.07 & -9.07 & 5.00 & 4.00 & -11.08 & 3.08 & 0.019 \\
\hline
Mean & staggered & & & 3.95 & -9.33 & 5.37 & 3.92 & -11.28 & 3.44 & \\
Std. dev. & staggered & & & 0.13 & 0.24 & 0.34 & 0.13 & 0.20 & 0.39 & \\
\hline
experimental & eclipsed & PP & All & 3.99 & -9.40 & 5.40 & 3.97 & -11.38 & 3.45 & 0.012 \\
experimental & eclipsed & PP & BHS & 4.03 & -9.46 & 5.42 & 4.00 & -11.45 & 3.45 & 0.013 \\
experimental & eclipsed & TZP & TM2 & 4.17 & -9.26 & 5.09 & 4.12 & -11.32 & 3.08 & 0.013 \\
frozen  & eclipsed & PP & All & 3.78 & -9.63 & 5.85 & 3.76 & -11.50 & 3.99 & 0.009 \\
frozen  & eclipsed & PP & BHS & 3.81 & -9.68 & 5.87 & 3.78 & -11.57 & 4.00 & 0.010 \\
frozen  & eclipsed & TZP & TM2 & 4.10 & -9.33 & 5.23 & 4.07 & -11.36 & 3.22 & 0.011 \\
relaxed & eclipsed & TZP & TM2 & 3.98 & -9.04 & 5.05 & 3.95 & -11.01 & 3.12 & 0.010 \\
\hline
Mean & eclipsed & & & 3.98 & -9.40 & 5.42 & 3.95 & -11.37 & 3.47 & \\
Std. dev. & eclipsed & & & 0.14 & 0.22 & 0.33 & 0.14 & 0.18 & 0.38 & \\
\hline
\end{tabular}
\caption{Calculated bare and renormalised parameters for the Hubbard model for ET monomers under various geometry relaxation schemes and with different pseudopotentials, basis sets and codes (see section \ref{sect:meth}). The `experimental' geometry is that reported for an ET molecule in  $\beta$-(ET)$_2$I$_3$,\cite{bI3:Leung} measured at  298 K, with the H atom (not observed in x-ray crystallography) positions relaxed. The `frozen' coordinate system was relaxed in the charge neutral state and held fixed for other charge states. The `relaxed' geometry was optimised at every charge state. All geometry relaxations were carried out in calculations using TZP basis functions and TM2 pseudopotentials; we have also carried out the relaxations in using the other methods in the table and find no significant differences. The abbreviation pseudo. (for pseudopotential) is used in this table and others below. The RMSE is taken from the fit to the classical Eqs (\ref{eqn:defnUf}). All values are in eV.}\label{tab:bI3}
\end{table*}

We present results for the experimental, frozen, and relaxed geometries in Table \ref{tab:bI3} using a variety of basis sets and pseudopotentials. The TZP basis sets consistently gives $U_m^{(v)}\sim0.2$ eV smaller than the PP basis sets; similar small differences are seen in $\epsilon_m$ and $\xi_m$. However, the trends between the different geometries are reproducible between the two calculations. These differences may be related to the fact that the PP orbitals are optimised for each charge state whereas the TZP numerical orbitals are the same for every charge state. This additional degree of freedom reduces the curvature of $E(q)$ which reduces \umvc and hence $U_m^{(v)}$. Whether pseudopotentials are used or all electron calculations are performed does not have a significant effect on the results. The experimentally measured\cite{Sato} gas phase ionisation energy of ET is 6.2 eV, in reasonable agreement with our calculated value of $\xi_m$. Note, however, that $\xi_{mc}$ is significantly smaller than the experimental value.

The trend between the different geometries is more interesting. Each time a geometry relaxation is performed \umv is reduced. This results form the increased number of degrees of freedom when more geometry relaxation is allowed. This can be seen from the plot of the energy for different charge states in Fig. \ref{fig:frozen vs relaxed}. For both of these curves both the geometry relaxation and SCF calculation were performed using the TZP basis sets and TM2 pseudopotentials. The two $q=0$ results are therefore, in fact, the same data point. As the charge is varied away from $q=0$ the geometry remains fixed in `frozen' calculation but relaxes in the `relaxed' calculation, thus the energy of `relaxed' data is lower than that of the `frozen'  calculation $q>0$. This effect gets larger as $q$ increases, thus \umv is smaller in the `relaxed' data than for the `frozen' calculation. The physical content of this result is simply the fact that intramolecular vibronic couplings act to lower \umvn.\cite{order1} However, one should expect this effect to be rather stronger {\it in vacuo} than in the crystalline environment where the motion of the molecule is significantly constrained. 

Thus we must face the question: should one use the `experimental', `frozen', or `relaxed' geometry to calculate \umvn? We believe that the `experimental' geometry gives the most useful information. Firstly, there are small differences in the reported geometries for different ET salts, and one would like to understand the effect of these on a single molecule before considering the effects of changes in the crystal structure on the emergent physics of the crystal. Secondly, the experiments effectively `integrate over' all of the relevant charge states and therefore provide an `average' conformation. Thirdly, the experiments naturally include the effects on the molecular conformation due to the crystalline environment, which are absent from {\it in vacuo} calculations. Therefore we now move on to consider the value of \umv found for the `experimental' geometry found in x-ray scattering experiments from a wide range of different \ET salts.

\section{Impurity scattering, polymorphism and chemical pressure} \label{sect:main}

We begin by studying polymorphism in (ET)$_2$I$_3$, for which the experimental literature contains more reported polymorphs than any other salt of ET. In Table \ref{tab:I3} we report the calculated values of the Hubbard parameters and the Coulomb integral for ET molecules in `experimental' geometries\cite{aI3:Bender,aI3:Emge,bI3:Leung,kI3:Kobayashi,tI3:Salameh} taken from the $\alpha$, $\beta$, $\kappa$, and $\theta$ polymorphs of (ET)$_2$I$_3$. All of the calculations in this section were performed using TZP basis sets and TM2 pseudopotentials. There is little variation in \umv across any of the polymorphs ($U_m^{(v)}=4.22\pm0.09$~eV). For the $\alpha$, $\kappa$, and $\theta$ polymorphs $\xi_m$ is extremely uniform ($\xi_m=4.840\pm0.008$~eV) but $\xi_m$ is somewhat larger in the $\beta$ polymorph. More interestingly $\xi_m$ is 0.12~eV larger for the staggered conformation of $\beta$-(ET)$_2$I$_3$ than it is for eclipsed conformation (the two conformations are sketched in Fig. \ref{fig:2D conformations}). To understand the significance of this result we must first briefly review a few experiments on \bI and a little of the theory of impurity scattering in unconventional superconductors.

\begin{table*}
\begin{tabular}{|l|l|l|l|l|l|l|l|l|l|l|}
\hline
Crystal type & conformation & $F_0$ & \umv & $\epsilon_m$ & $\xi_m$ & \umvc & $\epsilon_{m c}$ & $\xi_{m c}$ & RMSE \\
\hline
$\alpha$\cite{aI3:Bender} & staggered & 5.15 & 4.27 & -9.10 & 4.83 & 4.19 & -11.22 & 2.84 & 0.027 \\
$\alpha$\cite{aI3:Emge} & staggered & 5.14 & 4.29 & -9.13 & 4.84 & 4.20 & -11.25 & 2.85 & 0.030 \\
$\beta$\cite{bI3:Leung} & staggered & 5.35 & 4.08 & -9.28 & 5.21 & 4.07 & -11.31 & 3.17 & 0.013 \\
$\beta$\cite{bI3:Leung} & eclipsed & 5.40 & 4.17 & -9.26 & 5.09 & 4.12 & -11.32 & 3.08 & 0.013 \\
$\kappa$\cite{kI3:Kobayashi} & eclipsed & 5.22 & 4.31 & -9.16 & 4.85 & 4.22 & -11.29 & 2.85 & 0.030 \\
$\theta$\cite{tI3:Salameh} & eclipsed & 5.40 & 4.21 & -9.05 & 4.84 & 4.15 & -11.14 & 2.84 & 0.022 \\
\hline
Mean & & 5.28 & 4.22 & -9.16 & 4.94 & 4.16 & -11.26 & 2.94 & \\
Std. dev. & & 0.12 & 0.09 & 0.09 & 0.16 & 0.05 & 0.07 & 0.15 & \\
\hline
\end{tabular}
\caption{Calculated bare and renormalised parameters for the Hubbard model for an ET molecule at the `experimental' geometry observed at ambient temperature and pressure in various polymorphs of (ET)$_2$I$_3$. The changes in conformation due to the crystal packing structure do not have a large effect on the value of \umvn. The calculated $\xi_m$ is larger for the $\beta$ polymorph than the others explored, indeed $\xi_m$ varies by less than $1\%$ among the other polymorphs. Note that there is a significant difference between the values of $\xi_m$ in the eclipsed and staggered conformations in the $\beta$ phase. This is consistent with the effects of conformational disorder on \bIn.\cite{disorderBeta} The calculations use  TZP basis sets and TM2 pseudopotentials. All values are in eV.}\label{tab:I3}
\end{table*}

The application of hydrostatic pressure, $P$, to \bI has a dramatic effect on the superconducting critical temperature, $T_c$. At ambient pressure $T_c \sim 1.5$~K but when the applied pressure reaches $P\sim 1$~kbar a discontinuous increase in $T_c$ ($\sim$7~K) is observed. The low $T_c$ state ($P \lesssim 1$~kbar) is labelled the \bL phase and the high $T_c$ state ($P \gtrsim 1$~kbar) is labelled the \bH phase. When the pressure on the \bH phase is decreased the material does not return to the \bL phase but rather $T_c$ is seen to further increase. Below $T\sim 130$~K the resistivity of the \bH phase is found to undergo a discontinuous decrease while no such anomaly is found in the \bL phase.\cite{Ginodman} Incommensurate lattice fluctuations have been observed in the \bL phase but they are absent in the \bH phase below $T\sim130$~K.\cite{Emge} The incommensurate lattice fluctuations are stabilised by variations in the conformational ordering of the terminal ethylene groups of the \ET molecules and thus can only exist in the presence of disorder.\cite{Ravy} For a more detailed review of this phenomenology see Ref. \onlinecite{Ishiguro}. 

It is well known that non-magnetic disorder can lead to the suppression of $T_c$ in unconventional (`non-s-wave') superconductors, as the ET salts are believed to be.\cite{Powell:strongCorrelations,group2d} The details of this suppression are described by the Abrikosov-Gorkov equation,\cite{Larkin}
\begin{eqnarray}
\ln \left(\frac{T_{c0}}{T_{c}} \right) = \psi\left( \frac{1}{2} +
\frac{\hbar}{4\pi k_BT_{c}}\frac{1}{\tau} \right) - \psi\left(
\frac{1}{2} \right), \label{eqn:AG}
\end{eqnarray}
where $T_{c0}$ is the superconducting transition temperature of a
pure sample, $1/\tau$ is the quasiparticle scattering rate and
$\psi$ is the digamma function. If we consider only the contributions to $\tau$ from terminal ethylene group disorder then
\begin{eqnarray}
\frac{\hbar}{\tau}=N_s\pi D(E_F) |\Delta\xi_m|^2,
\end{eqnarray}
where, $N_s$ is the number of molecules in the staggered conformation, $D(E_F)$ is the density of states at the Fermi energy, 
and $\Delta\xi_m=0.12$ eV is the difference between the values of $\xi_m$ calculated in the staggered and eclipsed phases.\cite{footxivep} It has previously been shown\cite{disorderBeta} that over both the $\beta_H$ and $\beta_L$ phases of \bI the critical temperature scales with the residual resistivity, which is proportional to the contribution to $1/\tau$ from impurity scattering, as predicted by the Abrikosov-Gorkov equation.

Our finding that the staggered and eclipsed conformations have a small but important difference in $\xi_m$ shows that the disorder in the conformational degrees of freedom of the terminal ethylene groups of the ET molecules, required to stabilise the lattice fluctuations, is responsible for the suppression of $T_c$ in the $\beta_L$ phase. The terminal ethylene disorder increases the quasiparticle scattering rate and that this leads to the observed different $T_c$'s of the \bH and \bL phases. A simple calculation based on the derivation in Ref. \onlinecite{disorderBeta} shows that the calculated value of $\Delta\xi_m$ accounts for the observed differences between the critical temperatures and residual resistivities of the $\beta_H$ and $\beta_L$ phases with a few percent of the molecules staggered conformation.

\begin{figure}
	\centering 		\epsfig{figure=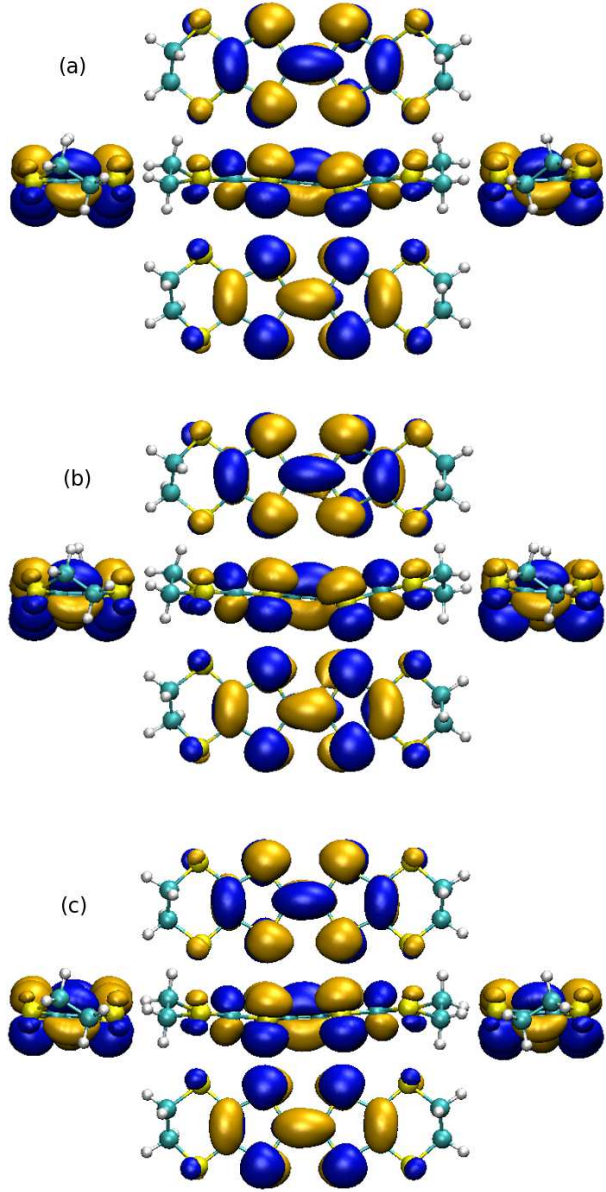, width=80mm, angle=0}
	\caption{(Color online) Orthographic projections of the isosurface of the HOMO of the ET molecule in `experimental' geometries taken from (a) eclipsed $\beta$-(ET)$_2$I$_3$, (b) staggered $\beta$-(ET)$_2$I$_3$, and (c) $\kappa$-(ET)$_2$Cu[N(CN)$_2$]Cl. Note  the great similarity of the HOMOs corresponding to the `experimental' geometries from eclipsed and staggered $\beta$-(ET)$_2$I$_3$ and the small electronic density on the terminal ethylene group, which is involved the change between the eclipsed and staggered conformations. Further, the  HOMOs of the ET molecule in the `experimental' geometries taken from \bI and $\kappa$-(ET)$_2$Cu[N(CN)$_2$]Cl are remarkably similar despite the fact that this geometry is taken from a different crystal polymorph with a different anion. This is consistent with our finding that the changes in the conformation of the ET molecule, in different polymorphs and in crystals with different anions, do not significantly affect \umvn. Colour indicates the sign of the Kohn-Sham orbital. All isosurfaces are $\pm$0.07 \AA$^{-3/2}$ and calculated in the charge-neutral state.  These calculations use  TZP basis sets and TM2 pseudopotentials. Animations showing different isosurfaces are available online.\cite{YouTube}} 	 \label{fig:iso}
\end{figure}

It is interesting that such significant differences are caused by the conformation changes in ET. In Fig. \ref{fig:iso}  we plot the Kohn-Sham orbitals corresponding to the HOMOs of eclipsed and staggered ET molecules at the `experimental' geometries found in \bIn. Note that there is very little electron density on the terminal ethylene groups, which are the only parts of the molecule in different positions in the `experimental' eclipsed and staggered phases. Furthermore, there are no significant changes between the HOMO electron densities on the other atoms between the two conformations.


%

%

\begin{figure}
	\centering
	 	\epsfig{figure=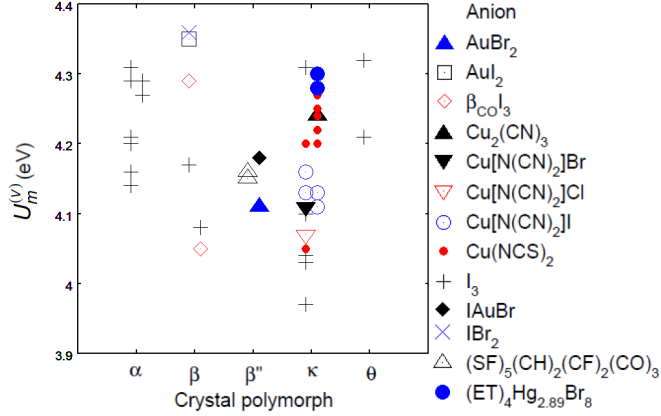, width=87mm, angle=0}
		\caption{(Color online) The effective Coulomb repulsion between electrons/holes, \umvn, on an ET monomer in the `experimental' geometries corresponding to different anions, conformations (eclipsed offset to the left, staggered to the right), temperatures, pressures, and crystal polymorphs. Note the limited range (3.9 -- 4.4 eV) of the ordinate. We see that \umv does not change significantly across the different (ET)$_2X$ crystals and has a mean value of $4.2 \pm 0.1$ eV.  The calculations use  TZP basis sets and TM2 pseudopotentials. Full details of the parameterisation are given in the supplementary information.} 	 \label{fig:manyUs}
\end{figure}

\begin{figure}
	\centering
		\epsfig{figure=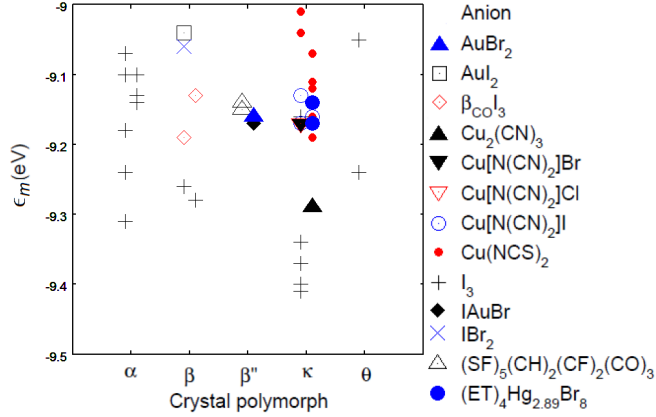, width=87mm, angle=0}
		\caption{(Color online) The site energy for electrons, $\epsilon_m$, on an ET monomer in the `experimental' geometries corresponding to different anions, conformations (eclipsed offset to the left, staggered to the right), temperatures, pressures, and crystal polymorphs.  The mean value is $\epsilon_m = -9.2 \pm 0.1$ eV.  The calculations use  TZP basis sets and TM2 pseudopotentials. Full details of the parameterisation are given in the supplementary information.} 	 \label{fig:manyes}
\end{figure}

\begin{figure}
	\centering
		\epsfig{figure=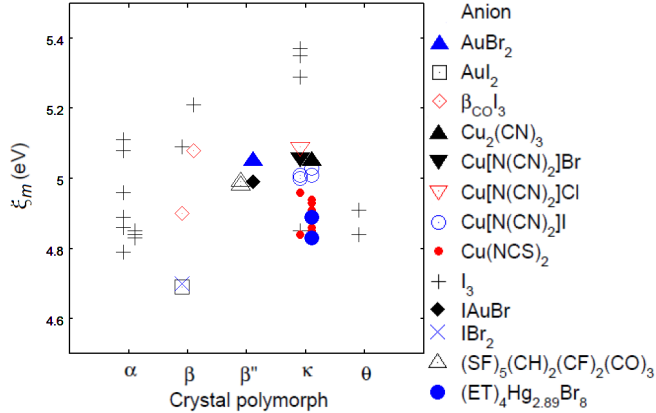, width=87mm, angle=0}
		\caption{(Color online) The site energy for holes, $\xi_m$, on an ET monomer in the `experimental' geometries corresponding to different anions, conformations (eclipsed offset to the left, staggered to the right), temperatures, pressures, and crystal polymorphs.  $\xi_m$ shows only slightly more variation than \umvn. However, small changes in $\xi_m$ are known to have significant effects on the superconducting state observed in ET salts.\cite{disorder,Analytis} (Because these ET molecules are quarter filled with holes in the salt, $\xi_m$, rather than $\epsilon_m$, is the relevant site energy to consider when discussing the role of disorder.) The mean value is $\xi_m=5.0\pm0.2$ eV.  The calculations use  TZP basis sets and TM2 pseudopotentials. Full details of the parameterisation are given in the suppmentary information.} 	 \label{fig:manyXis}
\end{figure}

It is well known that changing the anion in ET salts (which changes the unit cell volume) has a remarkably similar effect to applying a hydrostatic pressure.\cite{Powell:strongCorrelations} Therefore, changing the anion is often referred to as `chemical pressure'.\cite{Powell:strongCorrelations,Ishiguro} Both hydrostatic and chemical pressure have dramatic effects on the phase diagram of ET salts. This is typically understood in terms of a variation of Hubbard model parameters, e.g., the ratio $U/t$, where $t$ is the relevant hopping integral.\cite{Powell:strongCorrelations,JaimeRev} Therefore, we now consider the changes in the Hubbard parameters caused by changing the anion. Figs. \ref{fig:manyUs}, \ref{fig:manyes}, and  \ref{fig:manyXis} show the calculated values of, respectively, \umvn, $\epsilon_m$, and $\xi_m$ for `experimental' geometries taken from x-ray scattering experiments on crystals with a wide range of anions and several different polymorphs. The crystallographic data sets\cite{bI3:Leung,aI3:Bender,aI3:Emge,kI3:Kobayashi,tI3:Salameh,bAuI2:Wang,bIBr2:Williams,bI3:Madsen,bIAuBr:Ugawa,bAuBr2:Mori,kNCN2:Geiser,kCl3kbar:Schultz,kNCS2:Rahal,kNCS2:Schultz,kCN3:Geiser,kHg:Li,bSF5CH2CF5CO3:Geiser} used to determine the nuclear positions include data taken at a range of temperatures and under different pressures. Details of the calculated Hubbard parameters, and the crystallographic measurements on which the calculations are based are given in the supplementary information.

It can be seen from Fig. \ref{fig:manyUs} that the none of the changes to the crystal (anion, polymorphism, conformation, temperature, or hydrostatic pressure) have a significant effect on \umvn, indeed we find that for all of the structures we have studied $U_m^{(v)}=4.2 \pm 0.1$ eV (here, and below, the reported `error' is simply one standard deviation in the data and does not reflect systematic errors, particularly those arising from our approximate, density functional, calculation of the total molecular energies). 

An important question is: how large does variation of $U/t$ have to be in order to explain the experimental results? We can estimate this from strongly correlated theories of the Hubbard model. Both the resonating valence bond theory,\cite{RVB} and cluster extensions to dynamical mean-field theory\cite{cluster} suggest that $U/t$ is required to vary by perhaps a factor of two in order to explain the experimental phase diagram, although variational quantum Monte Carlo results suggest that the variation need not be quite this large.\cite{Liu}

Our results show that \umv is a transferable property and does not play any significant role in the `chemical pressure' effects observed in the behaviour of the ET salts. This means that either the chemical pressure comes from $\delta U_m^{(p)}$ and/or the elements of ${\cal H}_{ij}$, or else the simple Hubbard model description is not sufficient to describe the behaviour of the ET salts.

Figs.  \ref{fig:manyes} and  \ref{fig:manyXis} show that the variation of $\epsilon_m=-9.2\pm0.1$ eV and $\xi_m=5.0\pm0.2$ eV is not much larger than that in \umvn. However, it is known that the ET salts are very sensitive to disorder\cite{disorder,Analytis} therefore these subtle effects may lead to much more profound effects in the extended system than the variation in \umvn. Note also that while the Hubbard $U$ is the same for electrons and holes the site energies are rather different, as expected from the discussion in section \ref{sect:model}, when $U$ is large.

The similarity between the electronic states of ET molecules in the conformations found in different salts is further emphasised by fact that the Kohn-Sham orbital corresponding to the HOMO is essentially the same in all of the molecules we have studied. As an example of this we plot the electronic density of the HOMO for an ET molecule in the `experimental' geometry taken from a crystal of \kCl in Fig. \ref{fig:iso}. This can be seen to be remarkably similar to the equivalent plots for \bI (also shown in Fig. \ref{fig:iso}) despite the fact the polymorph and the anion are different.

%

\section{Conclusions}\label{sect:conc}

In summary, we have calculated the Hubbard $U$ and site energy for an isolated ET molecule. We found that $U_m^{(v)}=4.2 \pm0.1$ eV for `experimental' geometries taken from a broad range of conformations, polymorphs, anions, temperatures, and pressures. That is, \umv is essentially the same for all of the compounds studied. 

The dependence of the macroscopic behaviour of the ET salts on (hydrostatic and chemical) pressure is usually understood in terms of the variation of the Hubbard parameters with pressure. Strongly correlated theories of the Hubbard model\cite{RVB,cluster,Liu} suggest that $U/t$ is require to vary by a factor of $\sim$2 in order to explain the  experimentally observed chemical pressure effect. Therefore the pressure dependence must be contained in either the correction to \um from the crystalline environment, $\delta U_m^{(p)}$, or the intermolecular terms in the Hamiltonian, ${\cal H}_{ij}$. The renormalised value of \umv is significantly smaller than the bare value of the Coulomb integral, $F_0=5.2\pm0.1$ eV across the same set of geometries, emphasising the importance of using the renormalised value of \umvn.

The site energy (for holes) $\xi_m=5.0\pm0.2$ eV varies only a little more than \umv  across the same set of geometries. However, we have argued that this variation plays a key role in understanding the role of disorder in ET salts in general and in explaining the difference between the $\beta_L$ and $\beta_H$ phases of \bI in particular.

\section{Acknowledgements}
We thank Julian Gale his for assistance with the SIESTA suite of programs and Noel Hush, Anthony Jacko, Ross McKenzie, Seth Olsen, Mark Pederson, Jenny Riesz, Jeff Reimers, Elvis Shoko, Weitao Yang, and particularly, Laura Cano-Cort\'es and Jaime Merino for enlightening conversations. This work was supported by the Australian Research Council (ARC) under the Discovery scheme (project number DP0878523) and by a University of Queensland Early Career Research grant. BJP  was the recipient of  an ARC Queen Elizabeth II Fellowship (project number DP0878523). Numerical calculations were performed on the APAC national facility under a grant from the merit allocation scheme.




\begin{thebibliography}{99}

\bibitem{YangScience}  For a recent review, see A. J. Cohen, P. Mori-Sanchez, and W. T. Yang,   Science  {\bf321}, 792 (2008). 

\bibitem{Powell:strongCorrelations} For a recent review, see B. J. Powell and R. H. McKenzie, J. Phys.: Condens. Matter \textbf{18}, R827 (2006). 

\bibitem{JaimeRev} For a recent review, see H. Seo, J. Merino, H. Yoshioka, and M. Ogata, J. Phys. Soc. Japan  {\bf75}, 051009  (2006).


\bibitem{GunnarssonBook} O. Gunnarsson, \emph{Alkali-Doped Fullerides: Narrow-Band Solids with Unusual Properties} (World Scientific, Singapore, 2004). 

\bibitem{Brocks}
G. Brocks, J. van den Brink, and A. F. Morpurgo, Phys. Rev. Lett. {\bf93}, 146405 (2004).

\bibitem{Laura} L. Cano-Cort\'es, A. Dolfen, J. Merino, J. Behler, B. Delley, K. Reuter, and E. Koch, Eur. Phys. J. B {\bf56}, 173 (2007). \bibitem{other:Imamura} Y. Imamura, S. Ten-no, K. Yonemitsu, and Y. Tanimura, J. Chem. Phys. \textbf{111}, 5986 (1999). 

\bibitem{other:Fortunelli} A. Fortunelli and A. Painelli, J. Chem. Phys. \textbf{106} 8041 (1997); \textbf{106} 8051 (1997); Phys. Rev. B {\bf 55}, 16088 (1997). 
\bibitem{other:Ducasse} L. Ducasse, A. Fritsch, and F. Castet, Synth. Met. \textbf{85}, 1627 (1997). 
\bibitem{other:Castet} F. Castet, A. Fritsch, and L. Ducasse, Journal de Physique I. \textbf{6} 583 (1996).

\bibitem{vibronic} E. Demiralp, S. Dasgupta, and W. A. Goddard III, J. Am. Chem. Soc. {\bf 117}, 8154 (1995); J. Phys. Chem. A {\bf 101}, 1975 (1997); E. Demiralp and W. A. Goddard III, \ibid {\bf 102}, 2466 (1998); 
A. Girlando, M. Masino, G. Visentini, R. G. Della Valle, A. Brillante, and E. Venuti, Phys. Rev. B {\bf 62}, 14476 (2000); A. Girlando, M. Masino, A. Brillante, R. G. Della Valle, and E. Venuti, \ibid {\bf 62}, 14476 (2000); {\bf66} 100507(R) (2002). 


\bibitem{Ishiguro} T. Ishiguro, K. Yamaji and G. Saito, \textit{Organic Superconductors} (Springer Verlag, Heidelberg, 1998).

\bibitem{chem-rev}
See, for example, J. Yamada, H. Akutsu, H. Nishakawa, and K. Kikuchi, Chem. Rev. {\bf104}, 5057 (2004); J. M. Fabre, \ibid {\bf 104}, 5133 (2004); U. Geiser and J. A. Schlueter \ibid {\bf104} 5203 (2004).
 
 \bibitem{reviews}
 See, for example, H. Kino and H. Fukuyama, J. Phys. Soc. Japan {\bf 65}, 2158 (1996);
  K. Kanoda, Physica C {\bf282-287}, 299 (1997); Hyperfine Interact. {\bf104}, 235 (1997);
R. H. McKenzie, Comments Cond. Matt. Phys. \textbf{18}, 309 (1998);
 C. Hotta, J. Phys. Soc. Japan {\bf 72}, 840 (2003);
 H. Seo, C. Hotta, and H. Fukuyama, Chem. Rev. {\bf 104}, 5005 (2004).
 

\bibitem{order1} B. J. Powell, M. R. Pederson, and T. Baruah, arXiv:cond-mat/0510205.

\bibitem{Canadell}
See, for example, J. M., Williams, J. R. Ferraro, R. J.  Thorn, K. D. Carlson, 
U. Geiser, H. H. Wang, A. M. Kini, M.-H. Whangbo, \textit{Organic 
Superconductors (including Fullerenes): Synthesis, Structure, 
Properties and Theory} (Prentice Hall, New Jersey, 1992); E. Canadell, Chem. Mater. {\bf10}, 2770 (1998); A. Painelli, A. Girlando, A. Fortunelli, \prb {\bf64}, 054509 (2001).

\bibitem{Martin} R. L. Martin and J. P. Ritchie, Phys. Rev. B {\bf 48}, 4845 (1993).
\bibitem{Antropov} V. P. Antropov, O. Gunnarsson, and O. Jepsen, Phys. Rev. B {\bf 46}, 13647 (1992).
\bibitem{Quong} M. R. Pederson and A. A. Quong, Phys. Rev. B {\bf 46}, 13584 (1992).
\bibitem{Gunnarsson} O. Gunnarsson, Phys. Rev. B {\bf 41}, 514 (1990). 

\bibitem{wrongU}
U. R\"ossler, \emph{Solid State Theory} (Springer, Berlin, 2004); 
G. D. Mahan, \emph{Many-Particle Physics} (Kluwer Accademic, New York, 2000). 


\bibitem{FreedSimple} 
R. L. Graham and K. F. Freed, J. Chem. Phys. {\bf96}, 1304 (1992);
C. M. Martin and K. F. Freed, J. Chem. Phys. {\bf100}, 7454 (1994);
J. E. Stevens, K. F. Freed, F. Arendt, and R. L. Graham, J. Chem. Phys. {\bf101}, 4832 (1994);
J. P. Finley and K. F. Freed, J. Chem. Phys. {\bf102}, 1306 (1995);
J. E. Stevens, R. K. Chaudhuri, and K. F. Freed, J. Chem. Phys. {\bf105}, 8754 (1996);
R. K. Chaudhuri and K. F. Freed, J. Chem. Phys. {\bf119}, 5995 (2003); 
R. K. Chaudhuri and K. F. Freed, J. Chem. Phys. {\bf122}, 204111 (2005).

\bibitem{Janak} J. F. Janak, Phys. Rev. B \textbf{18}, 7165 (1978). 
\bibitem{GGA} D. C. Langreth and M. J. Mehl, Phys. Rev. B. \textbf{28}, 4 (1983). 
\bibitem{Perdew} 
J. P. Perdew and M. Levy, Phys. Rev. Lett. {\bf51}, 1884 (1983);
L. J. Sham and M. SchlŸter, Phys. Rev. Lett. {\bf51}, 1888 (1983).

\bibitem{PBE} J. P. Perdew, K. Burke, and M. Ernzerhof, Phys. Rev. Lett. \textbf{77}, 18 (1996). 
\bibitem{Yangsotherpapers}
A. J. Cohen, P. Mori-S\'anchez, and W. Yang, Phys. Rev. B {\bf77} 115123 (2008); Phys. Rev. Lett. {\bf100} 146401 (2008); arXiv:0809.5108

\bibitem{SIESTA} J. M. Soler, E. Artacho, J. D. Gale, A. Garc\'ia, J. Junquera, P. Ordej\'on, and D. S\'anchez-Portal, J. Phys.: Condens. Matter. \textbf{14}, 2745 (2002). 
\bibitem{NRLMOL} M. R. Pederson and K. A. Jackson, Phys. Rev. B. \textbf{41}, 7453 (1990); \ibid \textbf{43}, 7312 (1991); K. A. Jackson and M. R. Pederson, \ibid \textbf{42}, 3276 (1990); A. Briley, M. R. Pederson, K. A. Jackson, D. C. Patton, and D. V. Porezag, ibid. \textbf{58}, 1786 (1998). A. A. Quong, M. R. Pederson, and J. L. Feldman, Solid State Commun. \textbf{87}, 535 (1993); D. V. Porezag, Ph.D. thesis, Technische Universitat, 1997, http://archiv.tu-chemnitz.de/pub/1997/0025. 
\bibitem{PP} D. V. Porezag and M. R. Pederson, Phys. Rev. B \textbf{54}, 7830 (1996). 
\bibitem{Sankey} O. F. Sankey and D. J. Niklewski, Phys. Rev. B \textbf{40}, 3979 (1989). 
\bibitem{TM2} N. Troullier and J. L. Martins, Phys. Rev. B \textbf{43}, 1993 (1991). 


\bibitem{this:sup} See the supplementary information for pseudopotential parameters (available on request). 

\bibitem{BHS} G. B. Bachelet, D. R. Hamann, and M. Schl\"{u}ter, Phys. Rev. B. \textbf{26}, 4199 (1982). 

\bibitem{bI3:Leung} P. C. W. Leung, T. J. Emge, M. A. Beno, H. H. Wang, J. M. Williams, V. Petricek, and P. Coppens, J. Am. Chem. Soc. \textbf{107}, 22 (1985). 

\bibitem{aI3:Bender} K. Bender, I. Hennig, D. Schweitzer, K. Dietz, H. Endres, and H. J. Keller, Mol. Cryst. Liq. Cryst. \textbf{108}, 359 (1984). 
\bibitem{aI3:Emge} T. J. Emge, P. C. W. Leung, M. A. Beno, H. H. Wang, J. M. Williams, M. - H. Whangbo, and M. Evain, Mol. Cryst. Liq. Cryst. \textbf{138}, 393 (1986).
\bibitem{kI3:Kobayashi} A. Kobayashi, R. Kato, H. Kobayashi, S. Moriyama, Y. Nishio, K. Kajita, and W. Sasaki, Chem. Lett. \textbf{16}, 459 (1987).

\bibitem{tI3:Salameh} B. Salameh, A. Nothardt, E. Balthes, W. Schmidt, and D. Schweitzer, Phys. Rev. B \textbf{75}, 054509 (2007).

\bibitem{bAuI2:Wang} H. H. Wang, M. A. Beno, U. Geiser, M. A. Firestone, K. S. Webb, L. Nu\~{n}ez, G. W. Crabtree, K. D. Carlson, and J. M. Williams, Inorg. Chem. \textbf{24}, 2466 (1985). 






 

\bibitem{bIBr2:Williams} J. M. Williams, H. H. Wang,  M. A. Beno, T. J. Emge, L. M. Sowa, P. T. Copps, F. Behroozi, L. N. Hall, K. D. Carlson, and G. W. Crabtree, Inorg. Chem. \textbf{23}, 3841 (1984). 

\bibitem{bI3:Madsen} D. Madsen, M. Burghammer, S. Fiedler, and H. Muller, Acta Crystal., Sect. B: Struct. Sci. \textbf{55}, 601 (1999). \bibitem{bSF5CH2CF5CO3:Geiser} U. Geiser, J. A. Schlueter, H. H. Wang, A. M. Kini, J. M. Williams, P. P. Sche, H. I. Zakowicz, M. L. Van Zile, and J. D. Dudek, J. Am. Chem. Soc. \textbf{118}, 9996 (1996). 
\bibitem{bIAuBr:Ugawa} A. Ugawa, K. Yakushi, H. Kuroda, A. Kawamoto, and J. Tanaka, Chem. Lett. \textbf{15}, 1875 (1986). \bibitem{bAuBr2:Mori} T. Mori, F. Sakai, G. Saito, and H. Inokuchi, Chem. Lett. \textbf{15}, 1037 (1986). 
\bibitem{kNCN2:Geiser} U. Geiser, A. J. Schultz, H. H. Wang, D. M. Watkins, D. L. Stupka, and J. M. Williams, Physica C. 475 (1991). \bibitem{kCl3kbar:Schultz} A.J. Schultz, U. Geiser, H.H. Wang, and J.M. Williams, Physica C. 277 (1993). 
\bibitem{kNCS2:Rahal} M. Rahal, D. Chasseau, J. Gaulthier, L. Ducasse, M. Kurmoo, and P. Day, Acta Crystal., Sect. B: Struct. Sci. \textbf{53}, 159 (1997). 
\bibitem{kNCS2:Schultz} A. J. Schultz, M. A. Beno, U. Geiser, H. H. Wang, A. M. Kini, and J. M. Williams, J. Solid State Chem. \textbf{94}, 352 (1991). 
\bibitem{kCN3:Geiser} U. Geiser, H. H. Wang, K. D. Carlson, J. M. Williams, H. A. Charlier, Jr., J. E. Heindl, G. A. Yaconi, B. J. Love, and M. W. Lathrop, Inorg. Chem. \textbf{30}, 2586 (1991).

\bibitem{kHg:Li} R. Li, V. Petricek, G. Yang, and P. Coppens, Chem. Mater. \textbf{10}, 1521 (1998).

\bibitem{footgeo}
The atomic coordinates of the geometries studied in this paper are given in the supplementary information (available on request).
\bibitem{FreedBridge} K. F. Freed, Acc. Chem. Res. \textbf{16}, 137 (1983).

\bibitem{FreedSecond} S. Iwata and K. F. Freed, J. Chem. Phys. {\bf96}, 1304 (1992). 
\bibitem{footSIC} As we are presenting DFT calculations it is also interesting to consider the self interaction correction (SIC), which is absent in wavefunction methods. If the SIC per electron were constant then the SIC would cancel entirely when \umv is calculated from Eq. (\ref{eqn:defnU}). Of course this is only approximately true, but nevertheless suggest that the SIC is less important for \umv than for $F_0$. 

\bibitem{Sato}
N. Sato, G. Saito, and H. Inokuchi, Chem. Phys. {\bf76}, 79 (1983).
\bibitem{Ginodman} V. B. Ginodman, A. V. Gudenko, L. N. Zherikhina, V. N. Laukhin, E. B. Yagubskii, P. A. Kononovich, and I. F. Shegolev, Acta Poly. {\bf 39}, 533 (1988).
\bibitem{Emge} T.J. Emge, P.C.W. Leung, M.A. Beno, A.J. Schultz, H.H. Wang, L.M. Sowa, and J. M. Williams, Phys. Rev. B {\bf 30}, 6780 (1984).
\bibitem{Ravy} S. Ravy, J. P. Pouget, R. Moret and C. Lenoir, Phys.  Rev. B, {\bf 37}, 5113 (1988).

\bibitem{group2d}
B. J. Powell, J. Phys.: Condens. Matter {\bf18}, L575 (2006).

\bibitem{Larkin}
A. I. Larkin JETP Lett. {\bf 2}, 130 (1965).

\bibitem{footxivep} Note that as the ET salts are quarter filled with holes (three quarters filled with electrons) $\xi_m$, not $\epsilon_m$, is the relevant quantity here.

\bibitem{disorderBeta} B. J. Powell, J. Phys. IV (France) {\bf114}, 363 (2004).








\bibitem{RVB} 
B. J. Powell and R. H. McKenzie, Phys. Rev. Lett. {\bf 94}, 047004 (2005); Phys. Rev. Lett. {\bf98}, 027005 (2007); J. Y. Gan, Y. Chen, Z. B. Su, and F. C. Zhang, Phys. Rev. Lett. {\bf94}, 067005 (2005).

\bibitem{cluster} 
P. Sahebsara and D. S\'en\'echal, Phys. Rev. Lett. 97, 257004 (2006); B. Kyung and A.-M. S. Tremblay Phys. Rev. Lett. {\bf 97}, 046402 (2006).

\bibitem{Liu}
J. Liu, J. Schmalian, and N. Travedi, Phys. Rev. Lett. {\bf94}, 127003 (2005).


\bibitem{disorder} B.J. Powell and R.H. McKenzie, Phys. Rev. B {\bf69}, 024519 (2004).

\bibitem{Analytis}
J.G. Analytis, A. Ardavan, S.J. Blundell, R.L. Owen, E.F. Garman, C. Jeynes, and B.J. Powell,
Phys. Rev. Lett. {\bf96}, 177002 (2006).

\bibitem{other:Imamura:source} U. Geiser, A. J. Schultz, H. H. Wang, D. M. Watkins, D. L. Stupka, J. M. Williams, J. E. Schirber, D. L. Overmyer, D. Jung, J. J. Novoa, and M.-H. Whangbo, Physica C {\bf174}, 475 (1991). 


\bibitem{other:Fortunelli:source} H. H. Wang, K. D. Carlson, U. Geiser, A. M. Kini, A. J. Schultz, J. M. Williams, L. K. Montgomery, W. K. Kwok, U. Welp, K. G. Vandervoort, S. J. Boryschk, A. V. Strieby Crouch, J. M. Kommers, D. M. Watkins, J. E. Schirber, D. L. Overmyer, D. Jung, J. J. Novoa, and M.-H. Whangboo, Synth. Met. \textbf{42}, 1983 (1991).

\bibitem{other:Ducasse:source} A. J. Schultz, H. H. Wang, J. M. Williams, and A. Filhol, J. Am. Chem. Soc. \textbf{108}, 7853 (1986). 

\bibitem{YouTube} See http://www.youtube.com/watch?v=3K2kP8hWpZI http://www.youtube.com/watch?v=wIz1cRsSdEs and http://www.youtube.com/watch?v=bNzUBAS6AFM .
\end{thebibliography}
\end{document}